# Impact of Atmospheric and Model Physics Perturbations On a High-Resolution Ensemble Data Assimilation System of the Red Sea

Sivareddy Sanikommu[1], Habib Toye[1], Peng Zhan[1], Sabique Langodan[1], George Krokos[1], Omar Knio[1], and Ibrahim Hoteit[1,*]

1. King Abdullah University of Science and Technology (KAUST), Thuwal 23955-6900, Saudi Arabia

*Corresponding Author: ibrahim.hoteit@kaust.edu.sa

**Abstract**

The Ensemble Adjustment Kalman Filter (EAKF) of the Data Assimilation Research Testbed (DART) is implemented to assimilate observations of satellite sea surface temperature, altimeter sea surface height and in situ ocean temperature and salinity profiles into an eddy-resolving 4 km-Massachusetts Institute of Technology general circulation model (MITgcm) of the Red Sea. We investigate the impact of three different ensemble generation strategies (1) *Iexp* – uses ensemble of ocean states to initialize the model on 1$^{st}$ January, 2011 and inflates filter error covariance by 10%, (2) *IAexp* – adds ensemble of atmospheric forcing to *Iexp*, and (3) *IAPexp* – adds perturbed model physics to *IAexp*. The assimilation experiments are run for one year, starting from the same initial ensemble and assimilating data every three days.

Results demonstrate that the *Iexp* mainly improved the model outputs with respect to assimilation-free MITgcm run in the first few months, before showing signs of dynamical imbalances in the ocean estimates, particularly in the data-sparse subsurface layers. The *IAexp* yielded substantial improvements throughout the assimilation period with almost no signs of imbalances, including the subsurface layers. It further well preserved the model mesoscale features resulting in an improved forecasts for eddies, both in terms of intensity and location. Perturbing model physics in *IAPexp* slightly improved the forecast statistics and also the placement of basin-scale eddies. Increasing hydrographic coverage further improved the results of *IAPexp* compared to *IAexp* in the subsurface layers. Switching off multiplicative inflation in *IAexp* and *IAPexp* leads to further improvements, especially in the subsurface layers.

# 1. Introduction

A major component of ocean data assimilation systems is the background (forecast) error covariance, which spreads the observed information from one region to another in the model space (*Edwards et al.,* 2015; *Hoteit et al.,* 2018). Misrepresentation of the background error covariance may also lead to imbalanced ocean states, which could cause detrimental effects on the forecasts (*Bannister,* 2008a,b). Earlier studies (e.g. *Ravichandran et al.,* 2013; *Sivareddy,* 2015; *Xie and Zhu*, 2010; *Toye et al.,* 2017; *Waters et al.,* 2017; and the studies cited in the review paper of *Martin et al.,* 2015) have indicated that such a situation generally arises with static estimates of the background error covariances, mainly due to absence/spuriousness of cross-correlations between prognostic variables. For instance, *Ravichandran et al.* (2013) and *Waters et al.* (2017) suggested that 3D-VAR resulted in degradations of non-observed variables when cross-correlations between prognostic variables were not accounted for. *Xie and Zhu* (2010) and *Toye et al.* (2017) reported similar limitations with Ensemble Optimal Interpolation (EnOI) assimilation systems that use climatological ensembles for describing the background covariance. Ensemble Kalman filters (hereafter EnKF) provide a rigorous framework for the time-evolution of the background covariance, which should provide more balanced ocean analyses. The success of an EnKF largely depends on the appropriate initial selection and evolution of its ensemble. Many studies (e.g., *Lawson and Hansen,* 2004; *Leeuwenburg et al.,* 2005) have, however, argued that traditional EnKFs often suffer from the fast collapse of the ensemble spread. Although ad hoc inflation strategies could be used to increase the ensemble spread (e.g. *Anderson and Anderson*, 1999; *Hoteit et al.*, 2002; *Zhang et al.*, 2004; *Whitaker and Hamill*, 2012; *Bowler et al.,* 2017), they may limit the impact of the flow dependent statistics developed in the EnKF (see review paper *Houtekamer and Zhang*, 2016).

Several studies with ensemble atmospheric data assimilation systems suggested that describing the background errors with approaches based on multi parameter (e.g. *Bowler et al.,*

2008; *Murphy et al.,* 2011), multi physics (e.g. *Fujita et al.,* 2007; *Meng and Zhang,* 2007; *Houtekamer et al.,* 2009) and multi boundary conditions (*Torn et al.,* 2006) in which the ensemble members are integrated with different (perturbed) configurations of the dynamical model may help mitigating for the collapses of the ensemble spread, and generally leads to improved forecasts (see review paper *Houtekamer and Zhang,* 2016). Many ocean studies later followed, implementing ensemble ocean data assimilation systems with perturbed atmospheric forcing and showing important improvements in the forecasts (e.g. *Lisaeter et al.,* 2003; *Evenson,* 2004; *Wan et al.,* 2008; *Shu et al.,* 2011; *Sakov et al.,* 2012; *Karspect et al.,* 2013; *Penny et al.,* 2015; *Sivareddy et al.,* 2017, 2019). Other studies assessed the background/model error characteristics based on free model ensemble runs perturbing different parameters (*Brankart et al.,* 2015), boundary conditions (*Sandery et al.,* 2014), bathymetry (e.g. *Lima et al.,* 2019) etc. It has been later suggested in a few studies (to the best of our knowledge from *Luc and Barth*, 2015 and *Kwon et al.,* 2016) that combining perturbed atmospheric forcing and model physics leads to improved state estimates in ensemble ocean data assimilation systems. However, owing to the nonlinear nature of the ocean model equations, perturbations in the atmospheric forcing or model physics might also increase the errors in the ocean model if they are applied into a region not well covered by observations (*Sivareddy et al.,* 2017). Ensemble assimilation systems were also subject to stability issues (potential threat for dynamical balances in the analysis) when implementing inflation alone strategies with atmospheric/ocean models (e.g. *Anderson,* 2009; *Hoteit et al.,* 2013). The observational coverage in the ocean is generally sparse, particularly in the subsurface. In this context, it is important to assess whether the perturbation strategies result in improvements in ocean forecasts without disrupting dynamical balances, a pressing problem in ocean data assimilation, as discussed in the reviews of *Martin et al.* (2015) and *Hoteit et al.* (2018).

In this study, we implement perturbed (ensemble) atmospheric forcing and model physics strategies within a 4km-resolution Ensemble Adjustment Kalman Filter-Massachusetts Institute of Technology general circulation model (EAKF-MITgcm) assimilation system of the Red Sea (RS), and assess their potential to address the aforementioned issues of dynamical imbalances while improving the ocean forecasts. The RS is indeed among the most unexplored parts of the Indian Ocean despite its important contributions to the salinity budgets of the intermediate layers all along the length of the western parts of the Indian Ocean (e.g., *Beal et al.,* 2000). The RS surface circulation is governed by mesoscale eddies (e.g. *Zhan et al.*, 2014, 2018) that are largely modulated by changes in the overlying atmosphere (e.g. *Zhan et al.,* 2016, *Zhan et al.,* 2018). The intermediate layers of the southern RS are insulated from the local changes in the atmosphere, particularly during the active intrusion of relatively cold-fresh Gulf of Aden Intermediate Water (GAIW) mass (*Yao et al.,* 2014a, b). The RS is not adequately covered by global ocean reanalyses, probably due to its narrow width (maximum width of ~300 km) and historically sparse observation coverage. Hence the present study is also a step toward developing a state-of-the art ocean forecasting and reanalysis system for the RS.

The paper is organized as follows. Section 2 describes the model and the assimilation system, and the design of the assimilation experiments. Data utilized for evaluating the assimilation system outputs are presented in Section 3. Results from the multiplicative inflation alone assimilation experiment are presented in Section 4, in which we investigate the issues related to the system stability. Sections 5 and 6 present and discuss the results from the assimilation experiments employing the different perturbation strategies. Section 6 also analyzes and discusses the enhanced EAKF-MITgcm abilities with the ensemble atmosphere and perturbed model physics strategies. Section 7 summarizes the results and main conclusions of the study.

## 2. Description of the assimilation system and experiments

*2.1. Ocean model*

We use the MITgcm which solves the Navier-Stokes equations using the implicit free surface, Boussinesq and hydrostatic approximations (*Marshall et al.,* 1997). The model is configured for the domain 30°E-50°E and 10°N-30°N covering the whole RS, including the Gulf of Suez, the Gulf of Aqaba, and part of the Gulf of Aden where an open boundary connects it to the Arabian Sea. The model is implemented on Cartesian coordinates at an eddy-resolving horizontal resolution of 0.04° x 0.04° and 50 vertical layers, with 4m spacing in the surface and 300m near the bottom. The bathymetry is derived from the General Bathymetric Chart of the Ocean (GEBCO, available at http://www.gebco.net/data_and_products/gridded_bathymetry_data). Unless specified, the model uses direct space time $3^{rd}$ order scheme for tracer advection, harmonic viscosity with the coefficients of 30 $m^2$/s in the horizontal and $7x10^{-4}$ $m^2$/s in the vertical direction, implicit horizontal diffusion for both temperature and salinity, and the K-Profile Parameterization (KPP) scheme (*Large et al.,* 1994) for vertical mixing with a vertical diffusion coefficient of $10^{-5}$ $m^2$/s for both temperature and salinity. The open boundary conditions (OBCS) for temperature, salinity, and horizontal velocity are prescribed daily from the Global Ocean Reanalysis and Simulation data (GLORYS; *Parent et al.,* 2003) available at 1/12° horizontal grid. A sponge layer of 5 grid boxes with a relaxation period of 1-day is implemented for smooth incorporation of open ocean conditions through the eastern boundary. The normal velocities at the boundary are adjusted to match the volume flux of GLORYS, which is estimated from GLORYS Sea Surface Height (SSH) variations inside the model domain. The resulting inflow at the eastern boundary ensures consistency between our model and GLORYS basin-scale SSH. This is different from our earlier MITgcm setup, in which normal velocities on the eastern boundary were adjusted to enforce zero net inflow (e.g. *Yao et al.,* 2014a, 2014b; *Toye et al.,* 2017; *Gittings et al.,* 2018; *Zhan et al.,* 2018; see

Supplementary Material Section 2 for a comparison of the model results from the previous and new OBCS setup). The model was spun up for 31 years starting from 1979 to 2010 using the European Center for Medium Range Weather Forecast (ECMWF) reanalysis of atmospheric surface fluxes of radiation, momentum, freshwater sampled every 6-hour and available on a 75km x 75km grid (*Dee et al.,* 2011). The model simulations have been extensively validated for the RS by earlier studies (e.g. *Yao et al.,* 2014a, 2014b; *Toye et al.,* 2017; *Gittings et al.,* 2018; *Zhan et al.,* 2018). For comparison with the assimilation runs (as further discussed in the next section), we ran the same model configuration for the year 2011 using a 6-hourly 50km x 50km atmospheric ensemble mean, with initial conditions obtained from a spin up run. The atmospheric ensemble mean is extracted from ECMWF atmospheric ensemble as made available through The Observing System Research and Predictability Experiment (THORPEX) Interactive Grand Global Ensemble project (TIGGE, *Bougeault et al.,* 2010). We refer to this model free-run experiment without assimilation as *Fexp*.

## 2.2. Assimilation scheme

Available observations are assimilated using EAKF available in the DART-MITgcm (Data Assimilation Research Testbed coupled to MITgcm) package (*Hoteit et al.,* 2013, 2015) implemented for the RS by *Toye et al.* (2017). Here we implement DART-MITgcm with 50-members, assimilate data every 3 days, with localization in the horizontal (not in the vertical) direction using a radius of ~300 km , and a multiplicative inflation factor of 1.1, as suggested by *Toye et al.* (2017). The choice of 50-member ensemble is solely due to the size of the available atmospheric ensemble. We assimilate four different types of observations (see Section 2.3 for more detail on observations and the observation error variances), satellite-based sea surface temperature (SST) and SSH, and in situ temperature and salinity profiles. Errors associated with these observations are assumed uncorrelated, i.e., observational error covariance is diagonal. The assimilation experiments are conducted over a 1-year period in

2011, starting from January 1st, 2011. Unless otherwise stated, the initial ensemble for the assimilation experiments (described below) is generated by randomly selecting 50 ocean states from *Fexp* hindcasts corresponding to ±15 days from January 1st, the starting date of assimilation. The initial ensemble so obtained is then re-centered around the ocean state of *Fexp* corresponding to 1st January, 2011.

*2.3. Assimilated observations*

Observations from three different sources are assimilated, including SST data extracted from a level-4 in situ and advanced very high resolution radiometer infrared satellite SST blended daily product available on a 0.25°x0.25° grid (*Reynolds et al.,* 2007), along-track satellite level-3 merged altimeter filtered sea level anomalies (SLA; corrected for dynamic atmospheric, ocean tide, and long wavelength errors) from Copernicus Marine Environment Monitoring Service (CMEMS; *Mertz et al.,* 2017), and in situ temperature and salinity profiles made available by *Ingleby and Huddleston* (2007). While the SST observations are uniformly distributed, given the spatially complete level-4 gridded product, the altimeter SLA and in situ T/S observations are sparse in the RS. For instance, there are 5898 (~244) SLA observations (in situ temperature profiles) during the month of January, 2011 (the entire year 2011) spanning the whole model domain. No in situ T/S observations are available between August and December, 2011 and the salinity observations are even more sparse (only ~110 in situ salinity profiles in the entire year 2011).

The SLA observations are derived by subtracting 20 year (1993-2012) mean sea surface height ($\eta_{MSSH}^{Obs}$) from instantaneous sea surface height ($\eta_{SSH}^{Obs}$) observations (AVISO 2015; *Mertz et al.* 2017). Matching this $\eta_{MSSH}^{Obs}$ with the model based $\eta_{MSSH}^{model}$ is practically not possible because the altimeter measurements of $\eta_{SSH}^{Obs}$ are based on a reference ellipsoid and the model doesn't use any such ellipsoid (e.g. *Vidard et al.,* 2009). Since SLA represents the variable part of the $\eta_{SSH}^{Obs}$, one can only update $\eta_{SSH}^{model}$ based on the altimeter SLA, by simply adding the

$\eta_{MSSH}^{model}$ to the SLA measurements (e.g. *Vidard et al.,* 2009; *Costa and Tanajura,* 2015; *Zuo et al.,* 2019). This has the disadvantage of not reducing SSH biases, if any, in the climatological $\eta_{MSSH}^{model}$. In the present study, climatological biases in $\eta_{MSSH}^{model}$ are endeavored to keep minimal by averaging outputs between 2002 and 2016 (15 years) from a free model run forced with 5km-resolution atmospheric fluxes dynamically downscaled from 75km-resolution ERA-interim product (*Viswanadhapalli et al.,* 2016) and ocean boundary conditions obtained from altimeter assimilated global ocean reanalysis product (*Parent et al.,* 2003). The outputs of this run have been extensively validated in previous studies (e.g. *Gittings et al.,* 2018; *Zhan et al.,* 2018). Adding satellite along-track SLA to the $\eta_{MSSH}^{model}$ is only meaningful when temporal variations of $\eta_{SSH}^{model}$ are allowed through the ocean boundary conditions, one of the major improvements of the present study compared to the *Toye et al.* (2017) EAKF-MITgcm configuration (see model description in Section 2.1 and Section S2 of Supplementary Materials). This greatly improves the assimilation results for SSH as shown in the results presented in the Supplementary Materials (Section S2).

Observation error variances, which comprises errors due to instruments and unresolved scales and processes and interpolation, is an important element of the data assimilation system (e.g. *Hoteit et al.,* 2010; *Sivareddy et al.,* 2019). Temporally static and spatially homogeneous observational error variance values of $(0.04 \text{ m})^2$, $(0.5°C)^2$ and $(0.3\text{psu})^2$ are used for the satellite along-track SLA, the in situ T and S, respectively. These error variances for T and S, which are chosen in accordance with the suggested ranges of in situ observational errors by earlier assimilation studies (e.g., *Richman et al.,* 2005; *Forget and Wunsch,* 2007; *Oke and Sakov,* 2008; *Karspeck,* 2016), are intended to account the expected dominant errors from unresolved scales and processes (*Sivareddy et al.* 2019). The SLA observational error of $(0.04 \text{ m})^2$, which is slightly larger than the suggested altimeter accuracy (AVISO 2015; *Hoteit et al.,* 2002), is based on the sensitivity of our assimilation system to various choices of error variances, (0.04

m)$^2$, (0.07 m)$^2$, and (0.1 m)$^2$ (results not shown). The best-fit at such overall altimeter accuracy ranges may be explained to the native grid resolution of altimeter observations (~14km) which is close to the resolving scales of 4 km-MITgcm, essentially limitting the contribution of error from unresolved scales and process. Interpolation errors, due to the inevitable data gaps during dust events and satellite coverage, is the major source of observational error for the satellite blended level-4 SST observations. The 25km x 25km spatial resolution of the assimilated satellite blended SST observations in the present study is already close to the scales resolved by our 4km-MITgcm (*Pielke* 1984; *Grasso*, 2000), which allows us to discard the representation error in this particular application. The specified observational error variances for SST vary between (0.1°C)$^2$ and (0.6°C)$^2$ in accordance with the errors specified in the level-4 gridded SST product of *Reynolds et al.* (2007).

*2.4. Assimilation experiments*

Three main assimilation experiments were conducted: *Iexp*, *IAexp*, and *IAPexp* as outlined in Table 1. The ocean model configuration of *Iexp* is the same as *Fexp*, differing only in terms of assimilating observations in *Iexp* (using EAKF with multiplicative inflation) using 50-member ensemble of model forecasts integrated from perturbed initial conditions. Therefore, in terms of background error covariance, *Iexp* accounts for uncertainties in the initial conditions only and through inflation. *IAexp* is the same as *Iexp* except that it also accounts for uncertainties in the atmospheric forcing by driving each ensemble model run during the forecast step by a different atmospheric field extracted from the 50-member atmospheric ensemble forcing of the TIGGE project (*Bougeault et al.,* 2010). Figure 1 displays the spread of various atmospheric forcing parameters along the RS axis (indicated in Figure 2d). The spread in the downwelling shortwave radiation (DSW) is more pronounced over the southern RS (south of 22°N). It further exhibits marked seasonal variations with peak values (reaching 20 W/m$^2$) in July-August and troughs in February. These spatiotemporal variations are more or less similar for the other forcing parameters, except for rainfall. The rainfall spread is not significant, which is due to the negligible amount of rainfall received by the RS (e.g. *Dasari et al.,* 2017). The large ensemble spread for different variables during July-August, on the other hand, can be attributed to large variations in the atmospheric model during the strong Tokar jet, a southern Red Sea strong wind jet blows during July-August from the African continent through the Tokar Mountain Gap.

Ocean general circulation models rely on various physical parameterization schemes to account for the effects of unresolved scales of motion (*Brankart et al.,* 2013; *Jia et al.,* 2015; *Andrejczuk et al.,* 2015; *Zhu and Zhang*, 2019; also please refer the review of *Fox-Kemper et al.,* 2019). Such schemes depend on different parameters that need to be tuned according to the model configuration and domain of interest. These constitute another source for model uncertainties (as discussed in *Brankart et al.,* 2015 for example), in addition to those of the atmospheric forcing. *IAPexp* is designed to account for three sources of background error: initial conditions, model physics and atmospheric forcing. In *IAPexp*, each ensemble forecast model run is integrated with a set of model physics randomly selected from a predefined dictionary of model physics (here after MPD) at each assimilation cycle, meaning that a model run with a certain set of model physics in a given cycle is integrated with a different set of model physics in the next cycle. Table 2 summarizes the designed MPD for *IAPexp*. Three different categories of model physics are selected in the MPD: horizontal diffusion, horizontal viscosity, and vertical mixing. These three include different flavors of Gent-McWilliams/Redi sub-grid-scale eddy parameterization schemes for horizontal diffusion (*Redi*, 1982; *Gent and McWilliams*, 1990; *Gent et al.,* 1995; here after GMREDI): slope clipping of *Cox* (1987) (hereafter GMREDI-clipping), tapering scheme of *Danabasoglu and McWilliams* (1995) (hereafter GMREDI-dm95), and tapering scheme of *Large et al.* (1997) (hereafter GMREDI-ldd92) that uses a minimum diffusion coefficient of 100 m$^2$/s, and two other configurations, one with simple-explicit harmonic diffusion coefficient value of 100 m$^2$/s and another with implicit diffusion (same as in *Fexp*). The horizontal viscosity category in the MPD includes three different schemes: simple-harmonic with a value of 30 m$^2$/s (same as in *Fexp*), simple-bi-harmonic of *Holland* (1978) with a value of 10$^7$ m$^4$/s, and harmonic flavor of Smagorinsky/Leith (*Smagorinsky,* 1993; *Griffies and Hallberg,* 2000; hereafter SMAGLEITH-harmonic and SMAGLEITH-Biharmonic) scheme with a value of 30 m$^2$/s. The Smagorinsky and Leith coefficients are respectively set to 2.5 and 1.85 as suggested by *Griffies and Hallberg* (2000) and *Leith* (1996). For vertical mixing, four different schemes are included in MPD: the default nonlocal K-Profile Parameterization scheme of *Large et al.,* (1994) (hereafter KPP), the schemes of *Pacanowski and Philander* (1981) (hereafter PP81), *Mellor and Yamada* (1982) (hereafter MY82), and *Gasper et al.* (1990) (hereafter GGL90). The vertical diffusivity coefficient is the same in all these schemes. For all other coefficients, we used the default values as provided in the MITgcm. This MPD was designed after a careful examination of a large number MITgcm simulations using various options of model physics. The basic criterion in the selection process was to obtain a set of model physics that provides distinct, but not spurious forecasts around the forecast ensemble mean. More information on the selection process and the comparison of forecasts under the selection process is provided in the Supplementary Materials.

In addition to the above assimilation experiments we have conducted two more experiments *IAPcruiseexp* and *IAcruiseexp* to assess the impact of the observational coverage. These experiments are identical to *IAPexp* and *IAexp* but further assimilate CTD observations of temperature and salinity profiles collected in the RS between 15$^{th}$ September and 8$^{th}$ October, 2011 (as indicated in Figure 5). This dataset includes 206 profiles collected by a joint Woods Hole Oceanography Institute (WHOI) and King Abdullah University of Science and Technology (KAUST) cruise along the eastern part of the RS, with a horizontal spacing of

10km (*Zhai et al.*, 2015; hereafter WHOI/KAUST summer cruise). The two experiments are run starting 15th September, 2011 (starting period of the summer cruise data) till the end of the year using the analysis *IAPexp* and *IAexp* ensembles on that date as initial conditions, respectively. For straightforward comparison, we used the same perturbed model configurations of *IAPexp (IAexp)* in *IAPcruiseexp (IAcruiseexp)*.

## 3. Data used for evaluating the assimilation solution

Unless otherwise stated, we analyze daily averaged ocean forecasts as they result from the different experiments. For the evaluation of the subsurface features, CTD observations of temperature and salinity from WHOI/KAUST summer cruise are utilized. Root-Mean-Square-Differences (RMSD) of the assimilated solution (i.e. analysis snapshots and daily averaged forecasts) for SST and SSH are computed with respect to the corresponding assimilated observations. Spatially, SST is compared to a high-resolution daily averaged level-4 SST product from the Operational Sea Surface Temperature and Sea Ice Analysis (OSTIA; *Stark et al.*, 2007; *Donlon et al.*, 2012). OSTIA is generated on a 0.054° (~6 km) grid by combining SST data from various satellites and in situ observations using an Optimal Interpolation (OI) technique.

Multi-mission altimeter merged satellite level-4 gridded Absolute Dynamic Topography (ADT) provided by CMEMS (here after CMEMS-L4; *Mertz et al.*, 2017) is used for spatial evaluations of model simulated sea surface height (SSH). The ADT product is available on a 0.25° grid with temporal resolution of one day for the RS. The maximum formal mapping error of the ADT (provided along with the CMEMS-L4 ADT product) during the analysis period 1st January to 31st December, 2011 is estimated to be between 1.8 cm - 4 cm in the southern RS and reaches up to 6.7cm in the northern RS (not shown). In order to use it to evaluate the assimilated SSH solution, we adjust the CMEMS-L4 ADT by replacing its 15-

year average by $\eta_{MSSH}^{model}$, similar to the treatment of the assimilated along-track SLA data (see Section 2.3).

**4. Impact of assimilation with the multiplicative inflation-alone strategy**

The *Iexp* solution is first compared against that of *Fexp* to provide insights on the issues of assimilation with the *Iexp* strategy. Figure 2 displays biases and correlations between SST forecasts from *Fexp* and OSTIA (a and d), and *Iexp* and OSTIA (b and e). *Iexp* reduces SST biases by about 0.5°C over the RS and yields improved correlations in the northern RS by about 0.05, but slightly degrades those in the southern RS. Examining the time evolution of the RMSD (with respect to the assimilated SST observations) of SST forecasts in the RS suggests that *Iexp* SST-RMSDs are always smaller than those of *Fexp*, with improvements reaching up to 0.6°C during July, 2011 (Figure 3a). However, its time-variations are not stable, with strong seasonality similar to that of *Fexp*. The posterior SST-RMSDs also reveal similar results (Figure 3b), suggesting an inefficient use of SST observations in the *Iexp* assimilation strategy.

Figure 3c and 3d plot RMSD (with respect to assimilated SSH observations) time series of daily-averaged SSH forecasts and posterior-SSH from *Iexp* and *Fexp*. Unlike the SST results, no marked seasonality is observed in SSH-RMSDs of *Iexp*, and it is closer to that of the RMSDs in CMEMS-L4. The RMSDs fluctuate between 4-9cm in *Iexp* and between 2-15cm in *Fexp*, with the average RMSDs around 5cm in *Iexp* and 8cm in *Fexp*. The SSH-RMSDs in *Iexp* are smaller than those of *Fexp*, except for a short period around the first week of June and during the period August-November, where the SSH-RMSDs in *Iexp* are larger by ~2cm compared to those of *Fexp*. The larger SSH-RMSDs in *Iexp* during summer seem to be due to the spurious forecast error correlations in *Iexp*, as discussed in the subsequent paragraphs.

Figure 4 displays in situ observations of SST and sea surface salinity (SSS) from WHOI/KAUST summer cruise overlaid on spatial maps of SST and SSS, averaged over the cruise period of 15th August-10th October, 2011, from satellite merged product and model/assimilation experiments. The spatial patterns of SST are better described in *Iexp* compared to *Fexp*. In contrast with SST, SSS is poorly represented in *Iexp*. For example, while the spatial patterns of SSS in *Fexp* blends reasonably well with the SSS observations (Figure 4f), *Iexp* shows anomalous freshening, particularly in the southern RS (Figure 4g). This may have also spread to the SSS by horizontal advection through the western coast of the RS (*Yao et al., 2014b*).

We further examined subsurface temperature and salinity profiles during the WHOI/KAUST summer cruise. Figure 5 displays subsurface patterns of temperature and salinity from the cruise and the corresponding (in space and time) forecasts of temperature and salinity as resulting from *Fexp* and *Iexp*. *Fexp* shows cold biases in the upper layers of the northern basin and warm biases in the subsurface layers throughout the RS axis. The salinity structure is reasonably well captured in *Fexp*, except for saline biases between 18°N-20°N. Assimilation with *Iexp* improves the cold biases in the north, but increases the warm biases in the subsurface layers and further increases salinity errors. The subsurface layers below 150 m indeed show ad hoc features of anomalous warm and saline waters. *Iexp* simulates the intrusion of cold and fresh waters between 60-80 m south of 18°N, as indicated in the observations and *Fexp*, but the *Iexp* waters are anomalously colder and fresher than in the observations.

The pockets of anomalous ad hoc features in *Iexp*, such as those discussed above, can be related to long-range spurious correlations in the forecast ensemble (e.g., *Evenson, 2009; Sivareddy et al., 2017*), which migrate through the filter updates with the observations. Figure 6 displays the spatial patterns of temperature analysis increment after the filter update on 1$^{st}$ October, 2011 at various depths (2m, 50m, 180m, and 300m). As can be seen, the temperature increment in *Iexp* is almost close to zero at the surface but is larger in the subsurface layers. The spatial patterns of the increments in the subsurface layers appear noisy. For instance, the region surrounding 17.5°N at 50 m, where spurious cold waters are noticeable in the subsurface comparison plots, show large (up to 0.5°C) positive and negative increments. Similar scattered patterns can be seen at 180 m, where spurious features of temperature and salinity were noticed. As discussed earlier, there is no single subsurface observation available in the entire RS during September and October, and only surface observations were assimilated during this period. Hence the resulting increments in the subsurface layers must have been propagated through the correlations with the surface layer in the forecast ensemble. Figure 7 shows vertical correlations in the temperature ensemble at various locations. As can be seen from Figure 7, the correlation profile is indeed scattered in *Iexp,* indicating spurious correlations in the sub-surface layers. Further, as can be seen from Figure 8, the multiplicative inflation alone strategy shows negligible spread at the surface and unrealistically large spread in the deeper layers of the southern basin from August to the end of the simulation, in agreement with earlier studies (*Anderson* 2009; *Bowler et al.*, 2017) that showed spurious spread with a multiplicative inflation alone strategy in sparsely observed (atmospheric) regions. Statistically, such a spurious ensemble spread in the subsurface layers causes large corrections/increments even for small surface innovations, which may further explain the unrealistic subsurface features in *Iexp* with this strategy. Several approaches have been proposed to mitigate such spurious corrections: as for example, adaptive inflation (e.g. *Anderson,* 2009; *Miyoshi,* 2011), explicit balance operators (*Weaver et al.,* 2005). Directly accounting for model uncertainties may be a more straightforward approach to address this issue as already demonstrated in atmospheric data assimilation (e.g. *Fujita et al.,* 2007; *Bowler et al.,* 2008; *Houtekamer et al.,* 2009), as also suggested by the assimilation results presented in the next two sections.

**5. Impact of ensemble atmospheric forcing**

This section discusses the results of assimilation with ensemble atmospheric forcing, i.e. *IAexp*. As can be seen in Figure 8, *IAexp*, leads to a more reasonable ensemble temperature spread in the whole ocean column, and also increases the ensemble spread at the surface. As expected, the temporal evolution of the surface spread in the ocean (Figure 8) is similar to that in the atmospheric forcing (Figure 1). The atmospheric ensemble forcing increases the ensemble spread in the ocean at 50m too, owing to its influence on the mixed layer depth variations and the intrusion of Gulf of Aden Intermediate waters (*Sofianos and Johns*, 2007; *Yao et al.,* 2014; *Xie et al.,* 2019). The deeper layers (e.g. 150m) generally exhibit less spatiotemporal variations, and are less influenced by the perturbed atmospheric forcing, hence show less spread compared to surface layers. Unlike *Iexp*, *IAexp* yields relatively steady correlations (Figure 7) and updates (Figure 6) in the entire ocean column. For instance, at (38°E, 21°N), the vertical correlations of temperature between the surface and subsurface in the forecast ensemble of *Iexp* on 1$^{st}$ October, 2011 shows scattered patterns in the vertical correlations. *IAexp* on the other hand shows positive correlations in the upper layers and insignificant correlations in the deep layers (Figure 7). As a result, the temperature analysis increments are less noisy and more organized in *IAexp*, with larger increments in the surface (Figure 6). This also helps mitigating spurious spread in the deeper layers, which was prominent in *Iexp* (compare Figure 8c and 8g).

The ocean forecasts from *IAexp* show no anomalous features of salinity, both at surface and subsurface (Figure 5). The cold and saline biases in the upper 50 m in *Fexp* north of 23°N and the warm biases in the subsurface layers are significantly improved with *IAexp*. The deepening of isotherms at 26°N in the observations is reasonably reproduced by *IAexp*. The intrusion of subsurface cold and fresh waters in the southern latitudes are also better represented in *IAexp* compared to *Iexp* and *Fexp*. The spatial patterns of surface parameters in *IAexp*, including SSS, are more blended with the observations (Figure 4). In addition, the RMSDs of SST and SSH in *IAexp* are smaller than the interpolated products of OSTIA and CMEMS-L4, respectively, suggesting the efficient use of observations during assimilation. The RMSDs of SSH and SST are respectively consistently below 8cm and 0.8 °C throughout the study period with a reduced seasonality, and are significantly improved compared to *Fexp* and

*Iexp* (Figure 3). Note, however, that such a seasonality of SST-RMSDs may be also attributed to increased errors of satellite measurements in dust covered regions (prevalent during summer in the RS as suggested by *Ravi et al.,* 2018) as we notice similar seasonality in the observation-based interpolated product of OSTIA. The improvements in SST (SSH) forecasts of *IAexp*, with respect to *Iexp* and *Fexp*, reach 0.4°C and 1°C (2cm and 6cm), respectively, during July (October and April), 2011. The cold biases in the Gulf of Aden in *Iexp* are also significantly improved with *IAexp* (Figure 2). There are noticeable improvements in the spatial patterns of temporal variability of SST in *IAexp* compared to that of *Fexp* and *Iexp* (Figure not shown). The correlations along the southeastern coast of the RS further increase from 0.85 in *Fexp* to 0.9 in *IAexp* (Figure 2).

In order to further assess the influence of *IAexp* on the eddies, which are important elements of the RS circulation, we compare the spatial patterns of SSH in *IAexp* and *Fexp* with those of the satellite measurements. Figure 9 displays the spatial map of SSH from the different assimilation experiments and observations in three distinct regimes. The upper panels, corresponding to 12$^{th}$ February, 2011, showcase the cyclonic eddy of the northern RS (e.g. *Yao et al.,* 2014a), a permanent feature that is mainly influenced by thermohaline forcing (*Zhan et al.,* 2018). The middle panels, showing SSH in the southern RS on 11$^{th}$ August, 2011, correspond to the regime during which a dipole eddy feature, generated under the action of strong cross-basin Tokar winds, is prevalent (e.g. *Zhai and Bower*, 2013; Zhan et al., 2018). The bottom panels, corresponding to 10$^{th}$ July, 2011, showcase the typical anti-cyclonic eddy in the Gulf of Aden (GoA) that is influenced by various factors, including its advection from the adjacent Indian Ocean through instabilities in the Somali current and modification by local wind (e.g. *Al Saafani et al.,* 2007). One can see from Figure 9 that there are noticeable differences between CMEMS-L4 and along-track SSH observations in terms of their magnitudes, which can be attributed to the sparse altimeter coverage in the region. Nonetheless,

CMEMS-L4 still captures the aforementioned cyclonic eddy in the northern RS, dipole eddies in the central RS, and anti-cyclonic eddy in the GoA, consistant with earlier studies (*Yao et al.,* 2014a; *Zhai and Bower*, 2013; *Al Saafani et al.,* 2007). *Fexp* captures the anti-cyclonic eddy of the GoA reasonably well, but it overestimates the intensity of the cyclonic eddy in the northern RS. Although the anti-cyclonic eddy of the dipole features in the central RS is reasonably well simulated by *Fexp*, it underestimates the intensity of the cyclonic eddy. This may be attributed to the insufficient resolution of the atmospheric forcing to represent the cross-basin Tokar jet as suggested by earlier studies (e.g. *Clifford et al.,* 1997; *Zai and Bower*, 2013; *Bower and Farrar*, 2015). *IAexp* significantly improves the SSH biases and the placement of the eddies. Comparing the results with along-track SSH indicates that *IAexp* outperforms even the interpolated altimeter product, i.e. *CMEMS-L4*. For instance, compared to *Fexp and CMEMS-L4*, the SSH in *IAexp* is close to along-track SSH in all three regimes, particularly in the northern RS. The size and intensity of the anti-cyclonic eddy in the GoA are better represented by *IAexp* compared to *Fexp and CMEMS-L4*. Regarding dual eddies in the southern RS, *IAexp* improves the intensity of the anti-cyclonic eddy and captures the cyclonic eddy too, which was completely overlooked by *Fexp*.

*Toye et al.* (2017) argued that an increase in the model spread through seasonal ensemble optimal interpolation improved the analysis for the assimilated variables, but degraded the RS forecasts due to the disruption of dynamical balances. Such imbalances manifest themselves as spurious vertical velocities in the assimilation system outputs (e.g. *Anderson et al.,* 2000; *Hoteit et al.,* 2010; *Raghukumar et al.,* 2015; *Waters et al.,* 2017; *Park et al.,* 2018). To assess the dynamical balances in *IAexp*, which has better spreads and results in significant improvements in the tracer fields and eddy features, we examined the daily averaged forecasts of vertical velocities. Figure 10 displays the maximum vertical velocity in the ocean column (here onwards $|W(z)|_{max}$) along the RS axis as resulting from the model free-run and assimilation experiments. $|W(z)|_{max}$ in *Fexp* exhibits important spatio-temporal variability, with larger magnitudes in the regimes dominated by eddies (e.g. north RS during winter) and water-mass confluence zones (e.g. south RS), in agreement with earlier studies (e.g. *Pedro and Joaquin*, 2001). Compared to *Fexp*, the spatial extent of $|W(z)|_{max}$ is increased in *Iexp*, with magnitudes reaching 40 m/d, particularly between 18°N -20°N during August-December, the period during which spurious structures of subsurface temperatures and salinities were noticeable. The spatio-temporal structures and magnitudes of $|W(z)|_{max}$ in *IAexp*, on the other hand, show no striking differences between *Fexp* and *IAexp*, suggesting no significant dynamical imbalances.

## 6. Impact of perturbed model physics

Here, we examine whether incorporating another source of background error from perturbed physics, in addition to the ensemble atmospheric forcing and initial conditions, would provide further improvements in our Red Sea particular setting compared to *IAexp*. As can be seen in Figure 8, combining the three sources of background error led to a smoother and increased spread in *IAPexp,* although no significant changes are noticeable in the timing of peaks and lows. The increase in the spread is larger in the surface than in the subsurface, which is related to the larger sensitivity of perturbed internal physics to the zones of larger kinetic energy. The *IAPexp* strategy has also improved the spread in the northern parts of the RS compared to *IAexp*. As a result, although the structure of horizontal (Figure not shown) and vertical correlations in *IAexp* and *IAPexp* are quite similar, they differ mainly in terms of smoothness. The temperature correlations are more robust in *IAPexp,* compared to *IAexp,* irrespective of the chosen location (e.g. Figure 7). The spatial patterns of temperature increments in *IAPexp* are better organized owing to such smooth and steady correlations. The abrupt jumps in vertical velocities are further improved in *IAPexp* (Figure 10d), indicating more dynamically balanced ocean forecasts.

Comparing the SST forecasts and posterior in *IAexp* and *IAPexp* reveals that the latter results in some improvements, although not substantial, in terms of RMSDs and correlations (Figure 2, 3a, and 3b). The RMSD time series of SSH forecasts and posterior also indicate slightly better state estimates from *IAPexp* (Figure 3b and 3d). Visual comparisons of the eddy features suggest a slight improvement in the position of the basin-scale eddies in *IAPexp*. For instance, the cold core cyclonic eddy of the northern RS is placed closer to the along-track SSH observations in *IAPexp*. The dipole eddies, particularly the cyclonic eddy, in the southern RS are also better represented in terms of size and magnitude. The size of the anti-cyclonic eddy in the GoA is further reduced in *IAPexp* in closer agreement with along-track SSH observations.

Comparing the assimilated solutions with the WHOI/KAUST summer cruise T and S observations that were not assimilated here (but only in experiments *IAcruiseexp* and *IAPcruiseexp* discussed in the next paragraph), indicates that *IAPexp* improves the warm biases in the deeper (below 150m) layers. In contrast to the improvements at the surface and deeper layers, *IAPexp* appears to limit the improvements obtained with *IAexp* at the intermediate layers. For instance, in situ observations show intrusion of fresh and cold waters south of 20°N at around 50m depth (Figure 5b). *Fexp* managed to capture the intrusion of this water mass, although they are restricted to the area within 17°N (Figure 5c and 5d). While *IAexp* improves the extension of the intrusion (Figure 6g and 6h), *IAPexp* fails to simulate this intrusion (Figure 5i and 5j). Also, *IAPexp* seems to degrade the salinity forecasts in the northern RS (Figure 5j and Figure 4i), which could have been avoided by not using multiplicative inflation (see next paragraph). Careful examination of the ocean forecasts in *IAPexp* indicates that it diffuses the high-resolution spatial features, mostly mesoscale (size ~50km). For example, as can be seen from Figure 9g and 9j, the anti-cyclonic eddy close to the eastern coast of the northern RS, which was reasonably well represented in *IAexp*, is not well reproduced in *IAPexp*. It is worth mentioning here that the MPD used for the ensemble generation in *IAPexp* was prepared while making sure that the members didn't diverge too much from each other in terms of their model outputs (please see Supplementary Materials Section S1). Thus, the high resolution features cannot be lost altogether in any individual background model of the ensemble, but the placement and magnitude of the features can slightly vary around the mean. The ensemble mean can miss such high resolution features even if they are present in the individual members, as averaging the ensemble smoothens out spatial variations. Hence the diffusion of the above high-resolution features in *IAPexp* can be attributed to the typical representational issue with the ensemble average taken as the final estimate of an ensemble assimilation procedure. The high resolution features are therefore not lost in *IAPexp*, but are hidden within the ensemble space. And, compared to *IAexp*, the smoother ensemble mean forecasts (Figure 4 and Figure 9) in *IAPexp* is due to the larger ensemble spread resulting from the perturbed physics.

To investigate the impact of multiplicative inflation on *IAPexp* (*IAexp*), we conducted one more experiment *APexp* (*Aexp*), similar to *IAPexp* (*IAexp*), in which multiplicative inflation was not used. Comparing the RMSDs of SSH and SST of *APexp* with their counterparts of *IAPexp* indicate that discarding inflation improves the SST (Figure 13a) and SSH (Figure 13b), with the improvements reaching up to 0.3°C and 1cm, respectively, during summer. The salinity (Figure 13j) and temperature (Figure 13i) biases in the subsurface are also considerably improved in *APexp*. Similar improvements are noticeable in *Aexp* also, suggesting that the assimilation system may rely less on multiplicative inflation when background errors are properly accounted for.

To further investigate the impact of the perturbed atmospheric forcing in *IAPexp*, we conducted another experiment (*IPexp*) without using an ensemble of atmospheric forcing (just the ensemble mean) in IAPexp, so basically accounting for uncertainties in the initial conditions and the physics only. This generally resulted in a degradation of the assimilation solution, increasing the SST and SSH RMSDs by ~15-25% (and up to 0.2°C and 3 cm - Figure 3). The ensemble generation using perturbed physics still however provides better assimilation results than the multiplicative inflation alone strategy.

In order to provide insights on the enhanced capabilities of *IAPexp* strategy for assimilating hydrographic observations, we conducted additional experiments with *IAPexp* and *IAexp* assimilating the in situ temperature and salinity profiles from the WHOI/KAUST summer cruise (which were so far used for validation only). Figures 11 and 12 respectively show the spatial patterns of SSS and subsurface features of temperature and salinity from these two experiments, which we refer to as *IAcruiseexp* and *IAPcruiseexp*. Almost no improvements can be reported in *IAcruiseexp* compared to *IAexp*. In contrast, *IAPcruiseexp* yields large improvements in the salinity structure in the south. The SSS observations are now well blended in *IAPcruiseexp*, better than any other experiment, with improved representation of the fresh water intrusion and saline water advection in the southern and central-eastern parts of the RS, respectively (Figure 12h). The deepening of the isotherm at 26°N and warm biases in the sub-surface layers are further improved in *IAPcruiseexp* (Figure 12g). The larger improvements resulting from the *IAPexp* strategy with the assimilation of additional observations suggest its higher potential for areas with more observational coverage.

**7. Summary and conclusions**

Three different assimilation strategies, (1) multiplicative inflation alone: *Iexp*, (2) multiplicative inflation with ensemble atmospheric forcing: *IAexp*, and (3) multiplicative inflation with ensemble atmospheric forcing and perturbed model physics: *IAPexp*, are implemented and tested within a 50-member 4 km ocean ensemble EAKF-MITgcm data assimilation system of the Red Sea assimilating SST, SSH, and in situ temperature and salinity data. These were compared against the model solution without assimilation, *Fexp*. The relative impact of these strategies on the ocean forecasts was thoroughly examined. *Iexp* mostly improved the surface compared to *Fexp* owing to the homogeneous coverage of SST and SSH observations. It, however, shows abnormal vertical velocities and substantial degradations in the sparsely observed subsurface layers due to spurious vertical background error correlations.

Accounting for uncertainties in the atmospheric forcing by driving the ocean forecasts with an ensemble of ECMWF fields significantly improved the spatial and vertical correlations in the forecast ensembles, thereby helping to obtain noticeable improvements in the ocean forecasts. The warm and cold biases of SST in the southern and northern RS are significantly improved with *IAexp*. The improvements, in terms of RMSDs, reach 1°C and 0.2 psu in the temperature and salinity forecasts, respectively. Substantial improvements were obtained in the subsurface layers too, with less temperature biases compared to *Fexp*. The size and location of the eddies are further better captured in *IAexp*. In addition, the *IAexp* ocean state estimates show very limited dynamical imbalances, which has long been a desirable property for ocean data assimilation systems.

Accounting for another source of background errors through perturbed model physics, the *IAPexp* strategy, further improved the forecasts. This helped reducing the warm biases in the deeper layers and enhanced the placement of the basin-scale eddies. *IAPexp* also provided more dynamically balanced ocean forecasts, owing to more robust correlations in the forecast ensemble. Discarding the multiplicative inflation altogether yielded further improvements. The results also suggest that assimilating more hydrographic observations emphasizes more the positive impact of the *IAPexp* assimilation strategy.

Improved and balanced ocean forecasts from ensemble atmospheric forcing and model physics experiments is an important step toward the development of the first-ever ocean forecasting and reanalysis system for the Red Sea. A high resolution reanalysis for this region will be important not only for exploring the physical dynamics in this historically sparsely observed basin, but to also deepen our understanding of the regional biological processes.

**Acknowledgements**

This work was funded with the Office of Sponsored Research (OSR) at KAUST under the Virtual Red Sea Initiative (Grant #. REP/1/3268-01-01). All model and assimilation experiments were performed on the KAUST supercomputing facility Shaheen-II, Saudi Arabia. The DART-MITgcm ensemble ocean data assimilation system is implemented with ROCOTO workflow and we thank Christopher Harrop, NOAA, USA for making it available to us. The software installation support from Dr. Samuel Kortas, KAUST is greatfully acknowledged. The data used in the present study will be made available online, under the FAIR DATA standards of JGR-Oceans, upon acceptance of this manuscript for publication.

**Tables with captions**

**Table 1.** Summary of the conducted experiments.

| Experiment | Initial condition | Atmospheric Forcing | Model physics | Assimilated observations | Multiplicative Inflation |
|---|---|---|---|---|---|
| *Fexp* | Single member on 1st January, 2011 | Ensemble mean | Standard | None | NA |
| *Iexp* | 50-member ensemble on 1st January, 2011 | Ensemble mean | Standard | Reynolds-SST, Altimeter SSH, and in situ temperature and salinity as available from *Ingleby and Huddleston* (2007) | 1.1 |
| *IAexp* | 50-member ensemble on 1st January, 2011 | 50-member ensemble | Standard | Reynolds-SST, Altimeter SSH, and in situ temperature and salinity as available from *Ingleby and Huddleston* (2007) | 1.1 |
| *IAPexp* | 50-member ensemble on 1st January, 2011 | 50-member ensemble | Random | Reynolds-SST, Altimeter SSH, and in situ temperature and salinity as available from *Ingleby and Huddleston* (2007) | 1.1 |
| *IPexp* | 50-member ensemble on 1st January, 2011 | Ensemble mean | Random, but same as *IAPexp* | Reynolds-SST, Altimeter SSH, and in situ temperature and salinity as available from *Ingleby and Huddleston* (2007) | 1.1 |
| *IAcruiseexp* | 50-member ensemble on 1st September, 2011 from *IAexp* | 50-member ensemble | Standard | Reynolds-SST, Altimeter SSH, and in situ temperature and salinity as available from *Ingleby and Huddleston* (2007), and CTD observations collected during WHOI/KAUST cruise | 1.1 |
| *IAPcruiseexp* | 50-member ensemble on 1st September, 2011 from *IApexp* | 50-member ensemble | Random, but same as *IAPexp* | Reynolds-SST, Altimeter SSH, and in situ temperature and salinity as available from *Ingleby and Huddleston* (2007), and CTD observations collected during WHOI/KAUST cruise | 1.1 |
| *Aexp* | 50-member ensemble on 1$^{st}$ January, 2011 | 50-member ensemble | Standard | Reynolds-SST, Altimeter SSH, and in situ temperature and salinity as available from *Ingleby and Huddleston* (2007) | No inflation |
| *APexp* | 50-member ensemble on 1$^{st}$ January, 2011 | 50-member ensemble | Random but same as *IAPexp* | Reynolds-SST, Altimeter SSH, and in situ temperature and salinity as available from *Ingleby and Huddleston* (2007) | No inflation |

**Table 2.** Dictionary of model physics and associated coefficients considered in the present study. Coefficients of vertical mixing schemes vary according to the typical values set by MITgcm, unless otherwise stated. In the table, entries in italic indicate the standard scheme. Each ensemble member of the experiments that use perturbed physics selects a scheme randomly from each column.

| Horizontal Viscosity | Vertical Mixing | Horizontal diffusion |
|---|---|---|
| *Simple-Harmonic with viscosity coefficient 30 $m^2/s$* | *KPP* | *Implicit diffusion for temperature and salinity* |
| Simple-Bi-harmonic with viscosity coefficient $10^7$ $m^4/s$ | PP81 | Explicit coefficients of 100 $m^2/s$ for temperature and salinity |
| SMAGLEITH-Harmonic with viscocity coefficient 30 $m^2/s$, Smag coefficient 2.5 and Leith coefficient 1.85 | MY82 | GMREDI-clipping, with background diffusion set to 100 $m^2/s$ |
|  | GGL90 | GMREDI-dm95 with background diffusion set to 100 $m^2/s$ |
|  |  | GMREDI-ldd92 with background diffusion set to 100 $m^2/s$ |

**Figures with captions**

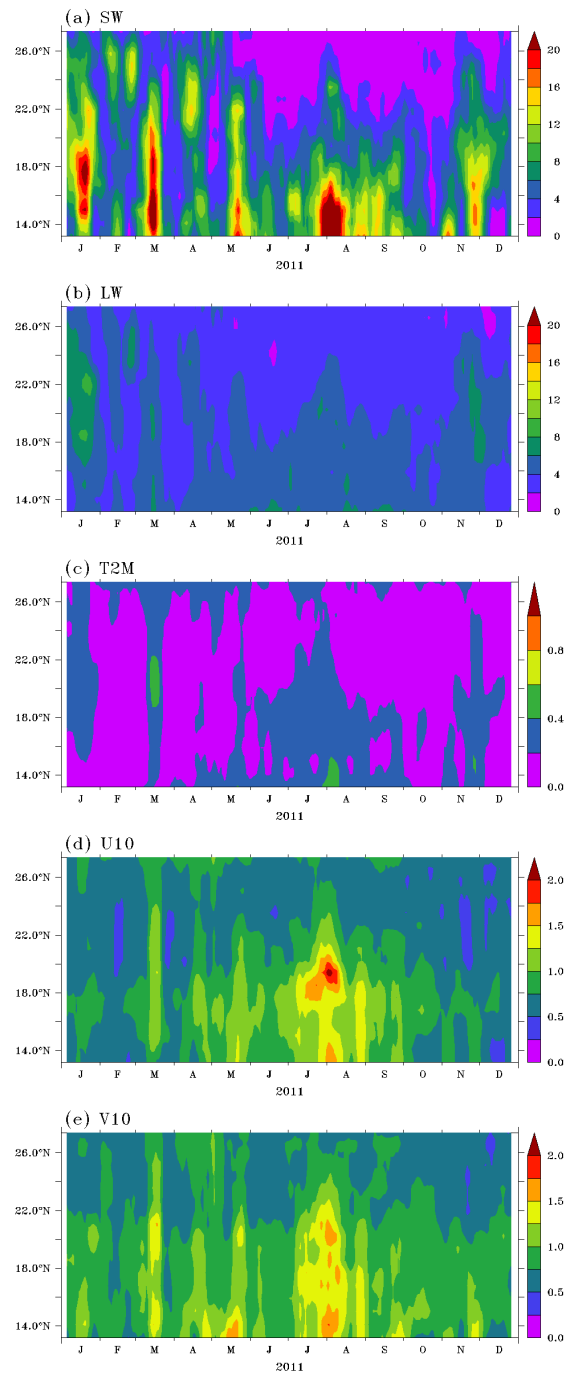

**Figure 1.** Temporal evolution of spread in the ECMWF ensemble atmospheric forcing along the axis of the Red Sea (as indicated in the first panel of Figure 2d). Spread is shown for (a) downwelling shortwave radiation (W/m$^2$), (b) downwelling longwave radiation (W/m$^2$), (c) air temperature at 2m (°C), (d) Zonal wind at 10m (m/s), and (e) Meridonal wind at 10m (m/s).

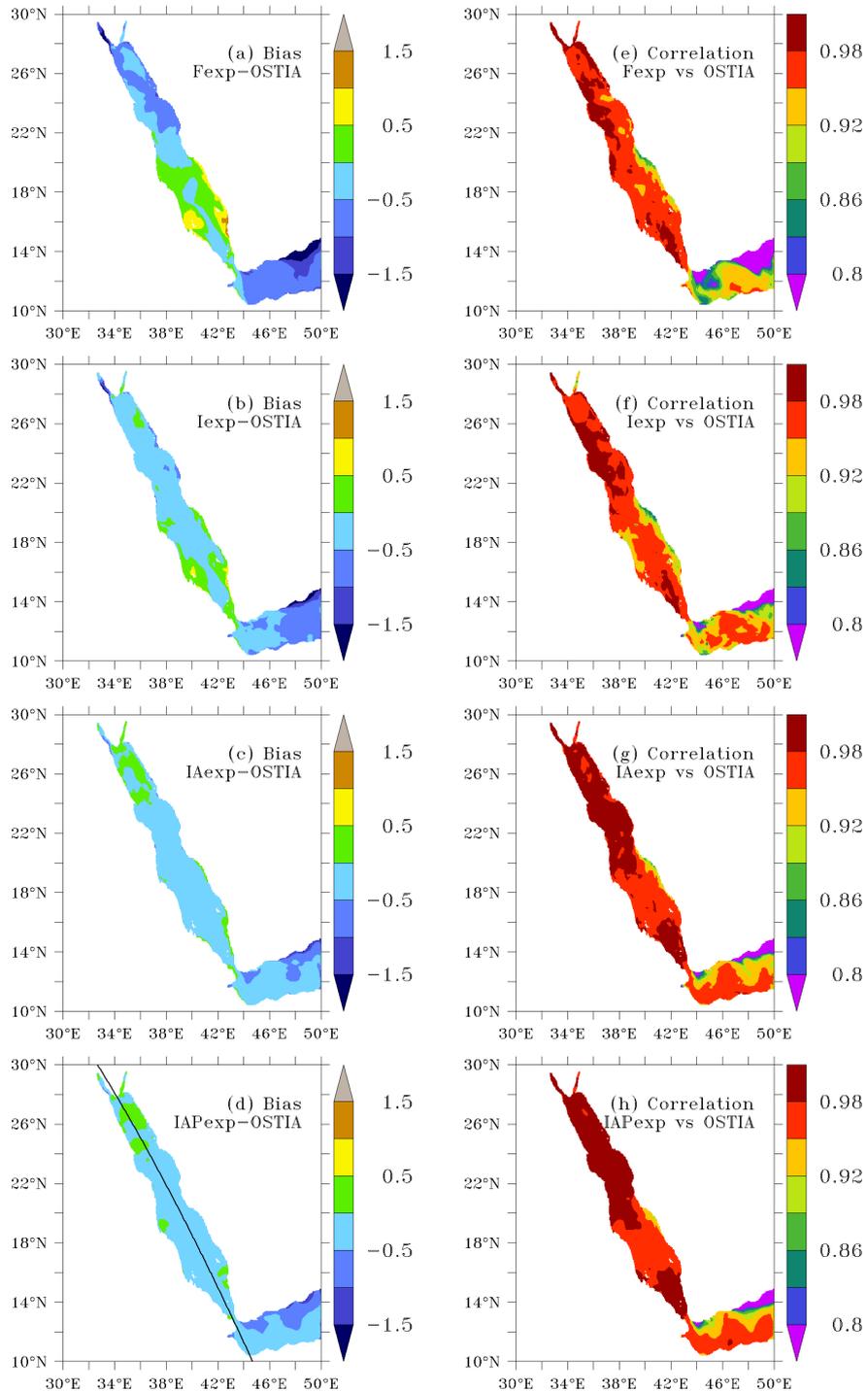

**Figure 2.** Spatial maps of SST bias (°C) between (a) *Fexp* and OSTIA, (b) *Iexp* and OSTIA, (c) *IAexp* and OSTIA, and (d) *IAPexp* and OSTIA. Panels (e-h) are the same as (a-d) except that they show correlations. Statistics are computed based on the period from 1$^{st}$ March, 2011 to 29$^{th}$ December, 2011. Black straight line in the panel 'd' is drawn to indicate the Red Sea axis, which is used in various plots of the study.

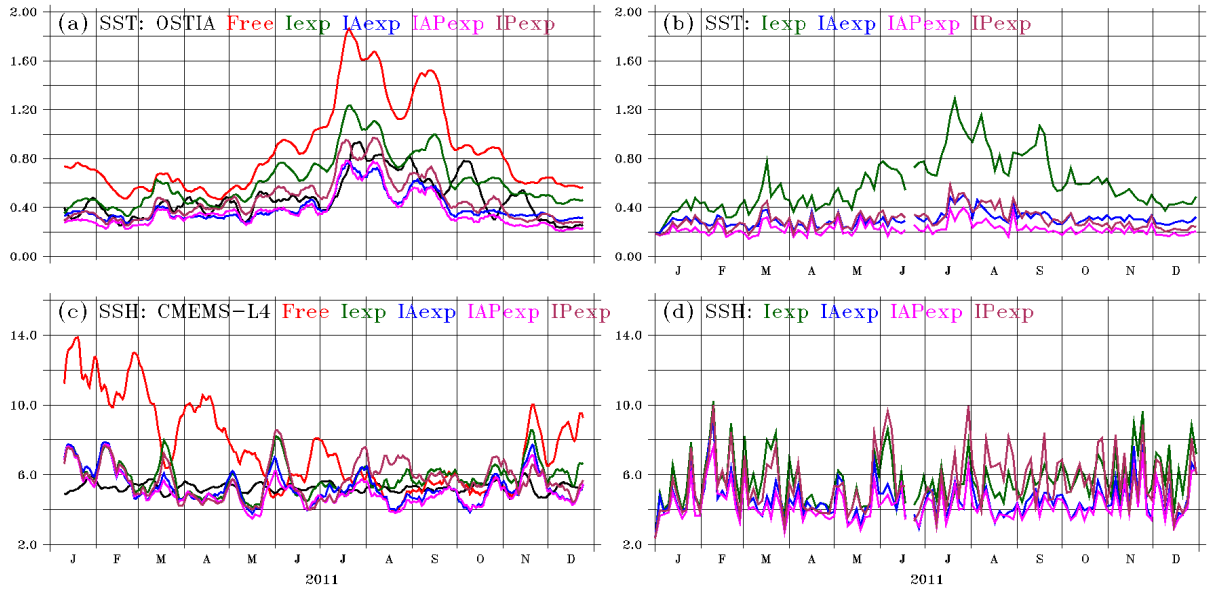

**Figure 3.** Time series of Root-Mean-Square-Difference (RMSD) for daily averaged (a) SST (c) SSH forecasts from level-4 gridded products (OSTIA for SST and CMEMS-L4 for SSH; black), *Fexp* (red), *Iexp* (green), *IAexp* (blue), *IAPexp* (pink), and *IPexp* (maroon). RMSD is computed by collocating the daily averaged model forecasts onto the corresponding observation locations. 10-day smoothing is applied to better highlight the differences between the assimilation results. Units are in "°C" and "cm" for SST and SSH, respectively. Panels (b) and (d) are similar to (a) and (c) but calculated for the analysis with no smoothing, as the plotting interval is already 3 days due to assimilation cycle.

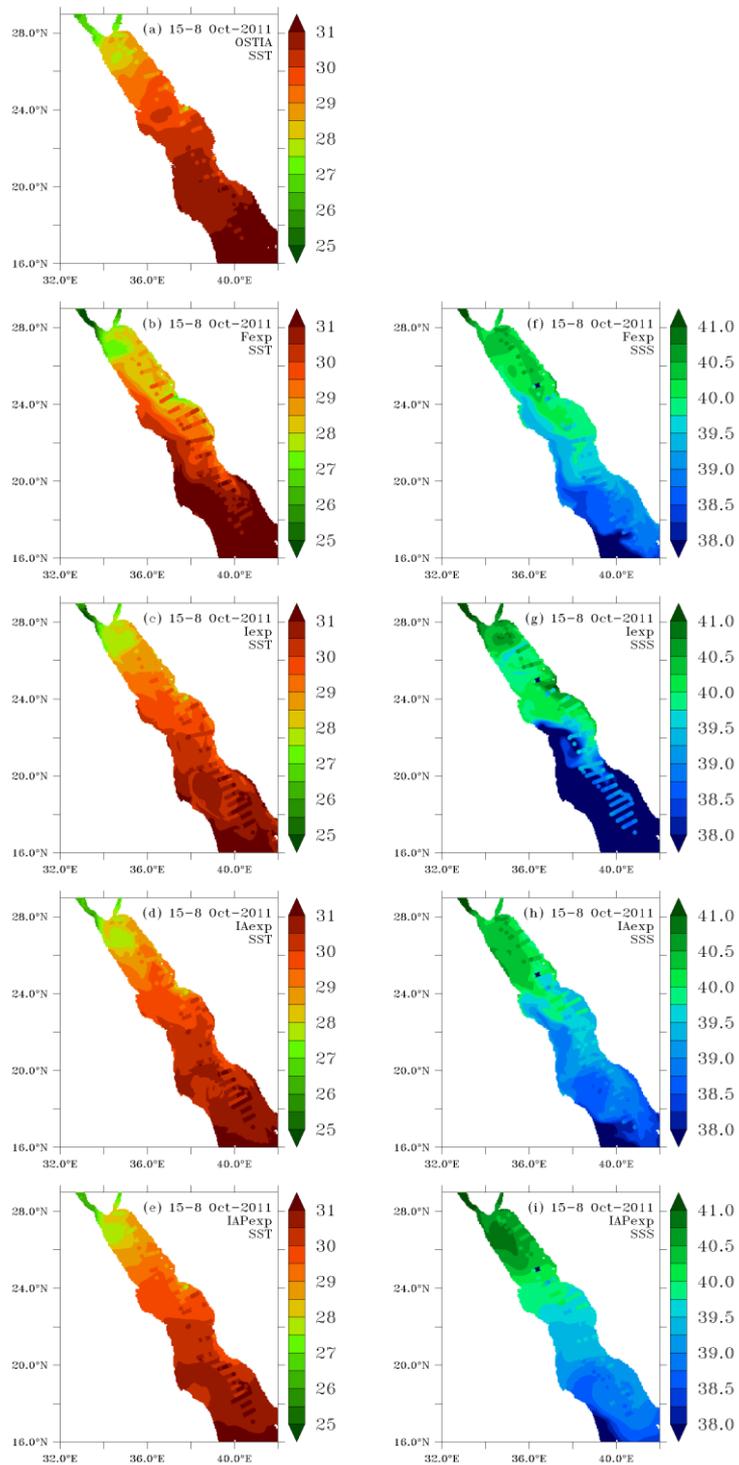

**Figure 4.** Spatial maps of temporally averaged SST (°C) from (a) OSTIA, (b) *Fexp*, (c) *Iexp*, (d) *IAexp*, and (e) *IAPexp* during the period pertained to the WHOI/KAUST summer cruise (15$^{th}$ September -8$^{th}$ October, 2011). Near surface in situ temperature from the CTD data collected during the summer cruise is also shown with filled circles on each plot. Panels (f-i) are same as (b-e) but for SSS in psu.

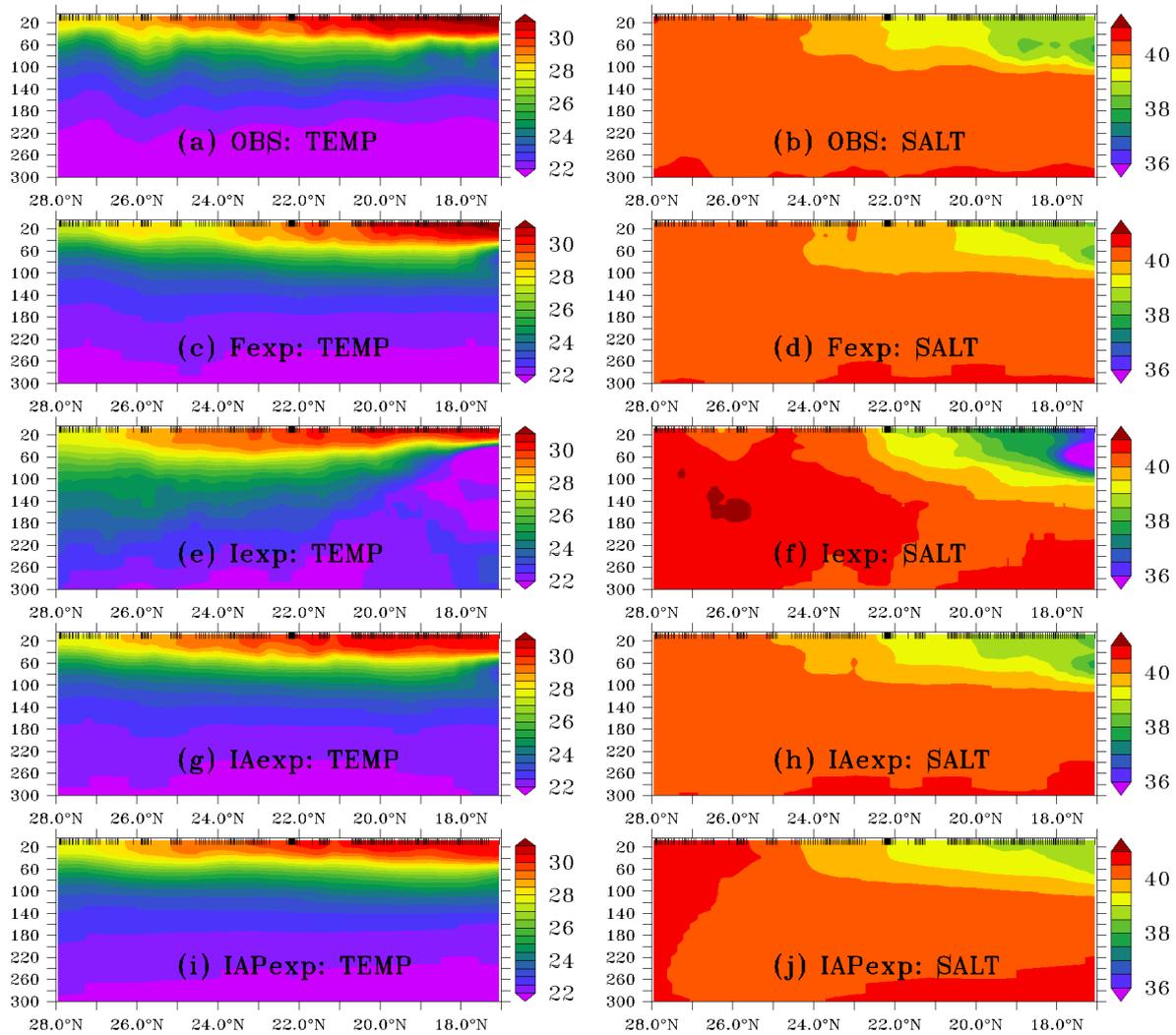

**Figure 5.** Subsurface temperature (in °C) and salinity (in psu) from in situ CTD observations (a-b) and the collocated (in space and time, during the WHOI/KAUST summer cruise conducted during 15th September – 8th October, 2011) daily averaged temperature and salinity forecasts as resulted from *Fexp* (c-d), *Iexp* (e-f), *IAexp* (g-h), and *IAPexp* (i-j). Temperature and salinity are smoothed by 1° and 10m in latitudinal and vertical direction to better highlight subsurface features. Latitudes corresponding to the cruise observation locations are indicated as black vertical dashes at the top of each panel.

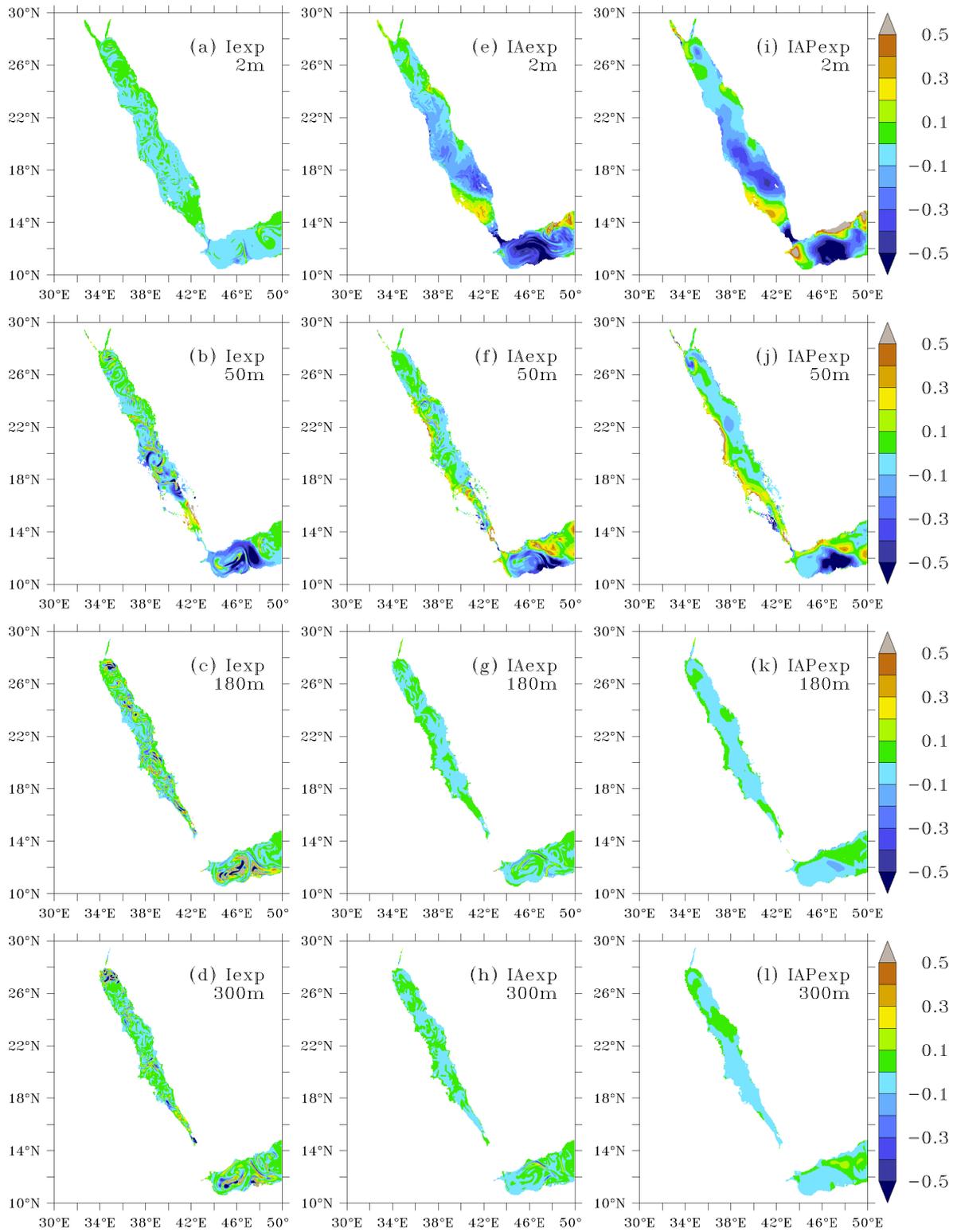

**Figure 6.** Ensemble mean temperature increments (analysis – forecast) in *Iexp* on 1$^{st}$ October, 2011 at (a) surface, (b) 50m, (c) 180m, and (d) 300m. Panels (e-h) and (i-l) show similar plots for *IAexp* and *IAPexp,* respectively. Units are in °C.

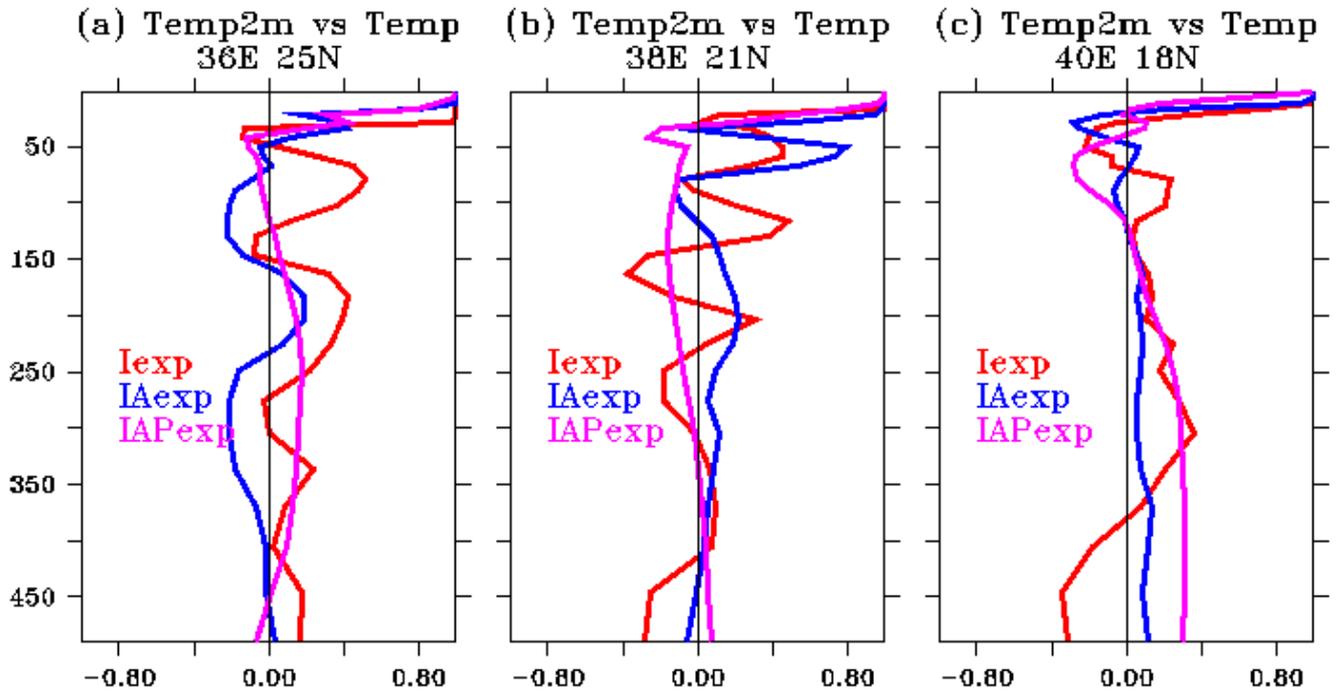

**Figure 7.** Zero-lag single point ensemble correlations in the vertical direction on 1st October, 2011 from *Iexp* (green), *IAexp* (blue), and *IAPexp* (pink), between SST and temperature at (a) North RS (36°E & 25°N), (b) central RS (38°E & 21°N) and (c) south RS (40°E & 18°N) locations.

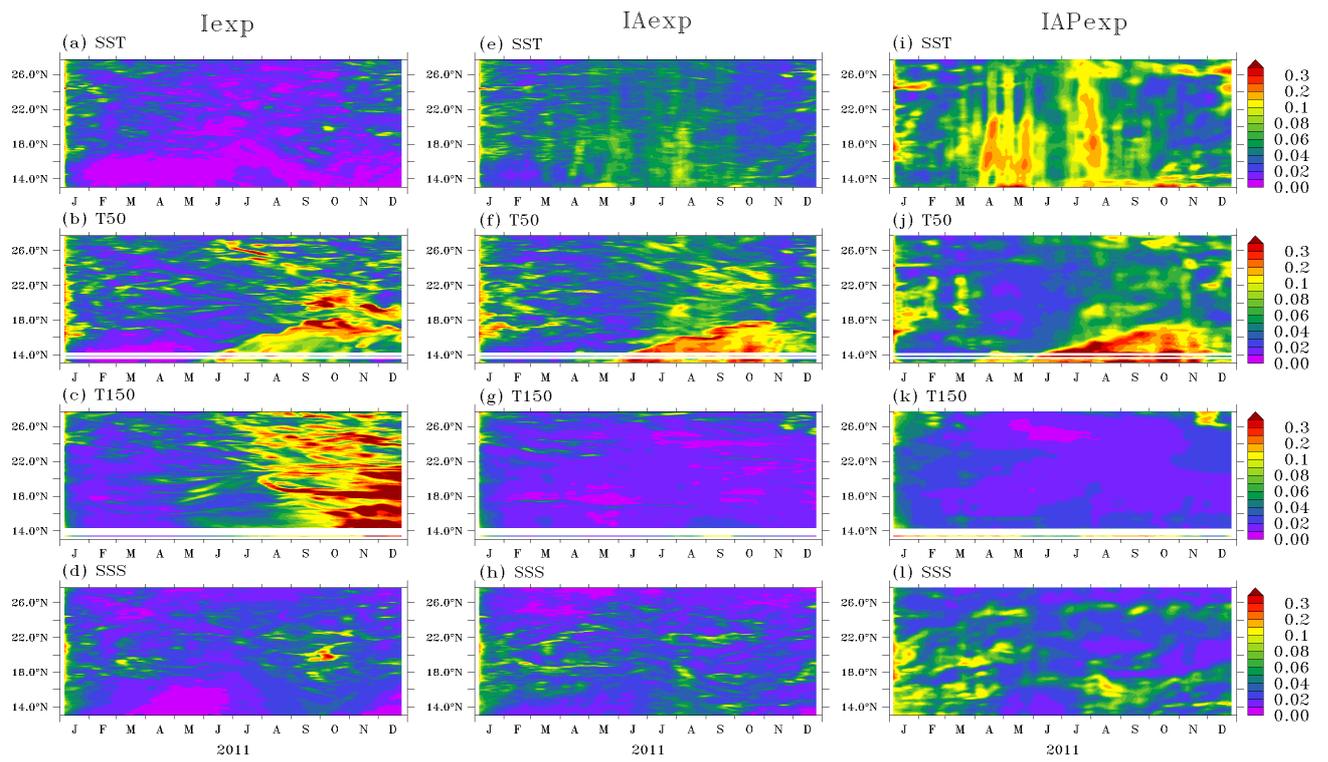

**Figure 8.** Temporal evolution of spread in various variables along the axis of the Red sea. Spread is shown for temperature (in °C) at surface, 50m, 150m, and sea surface salinity (in psu) from *Iexp (a-d)*, *IAexp (e-h)*, and *IAPexp (i-l)*.

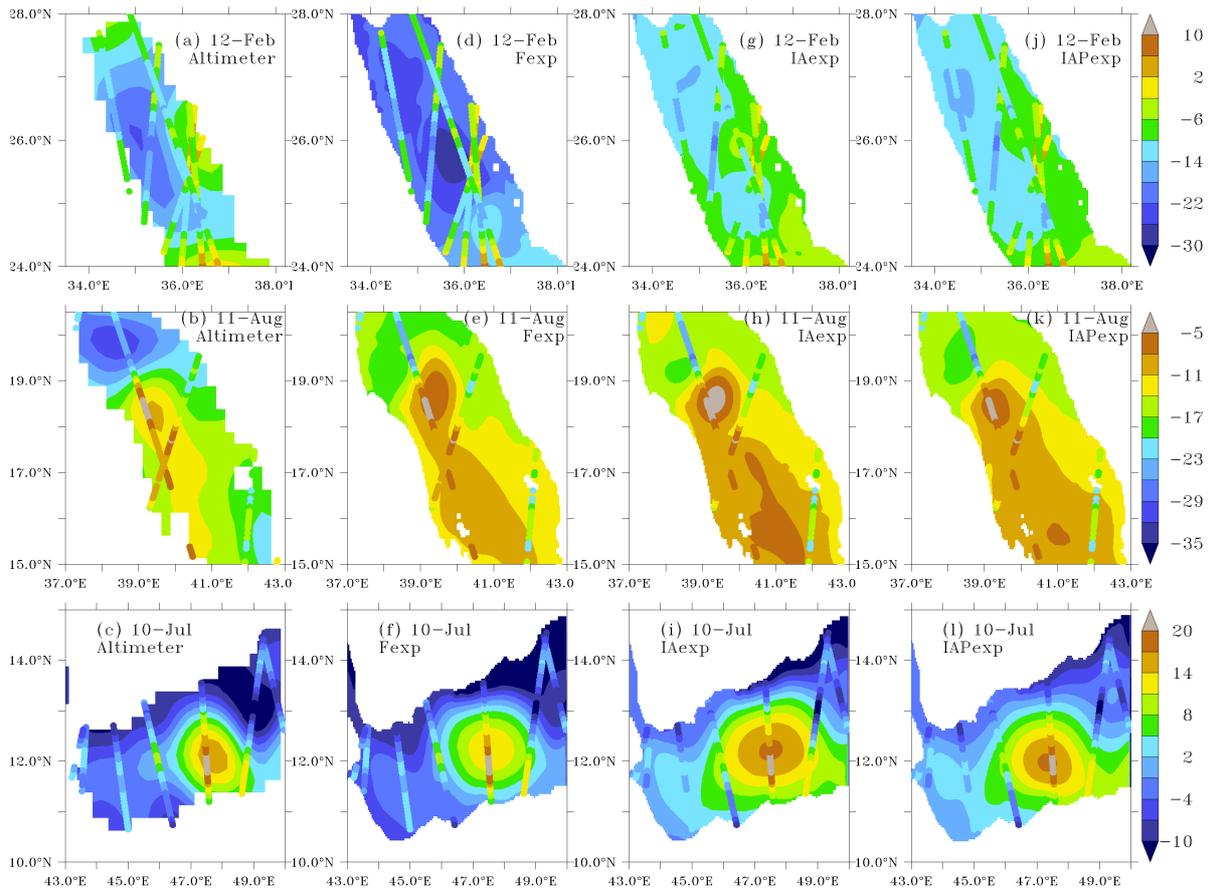

**Figure 9.** Spatial maps of 3-day averaged SSH forecast (in cm) corresponding to 12$^{th}$ February, 2011 (top), 11$^{th}$ August, 2011 (middle), and 10$^{th}$ July, 2011 (bottom) from (a-c) merged altimeter CMEMS-L4. Panels (d-f), (g-i), and (j-l) show similar plots from *Fexp*, *IAexp* and *IAPexp* forecasts, respectively. Along track SSH observations are also overlaid on each map.

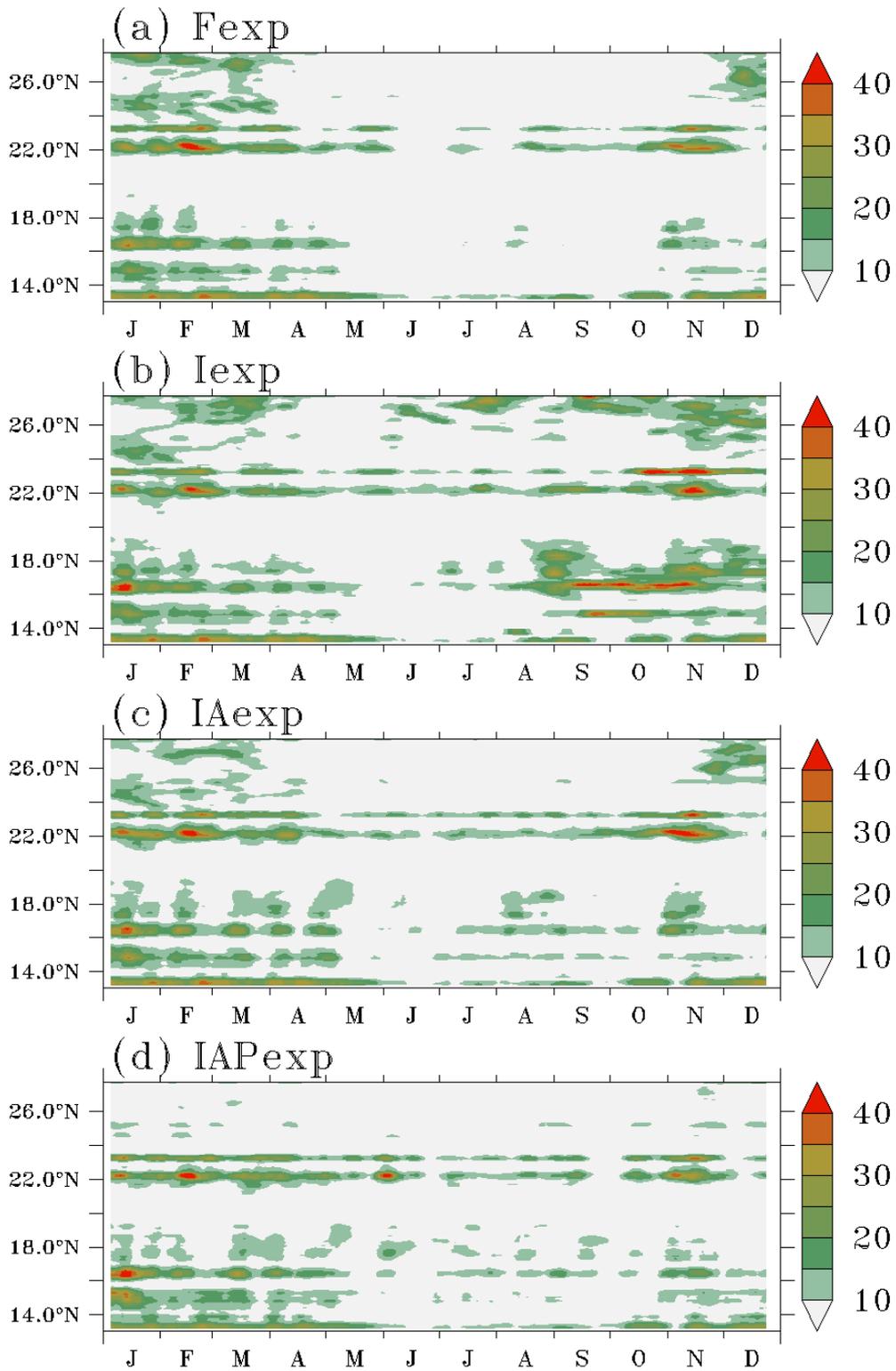

**Figure 10.** Temporal evolution of the daily averaged forecasts of vertical velocity $|W(z)|_{max}$(m/day) in the ocean column along the axis of the Red Sea from (a) *Fexp*, (b) *Iexp*, (c) *IAexp*, and (d) *IAPexp*. $|W(z)|_{max}$(m/day) is smoothed by 0.2° and 5-days in the latitudinal and temporal direction to better highlight the features.

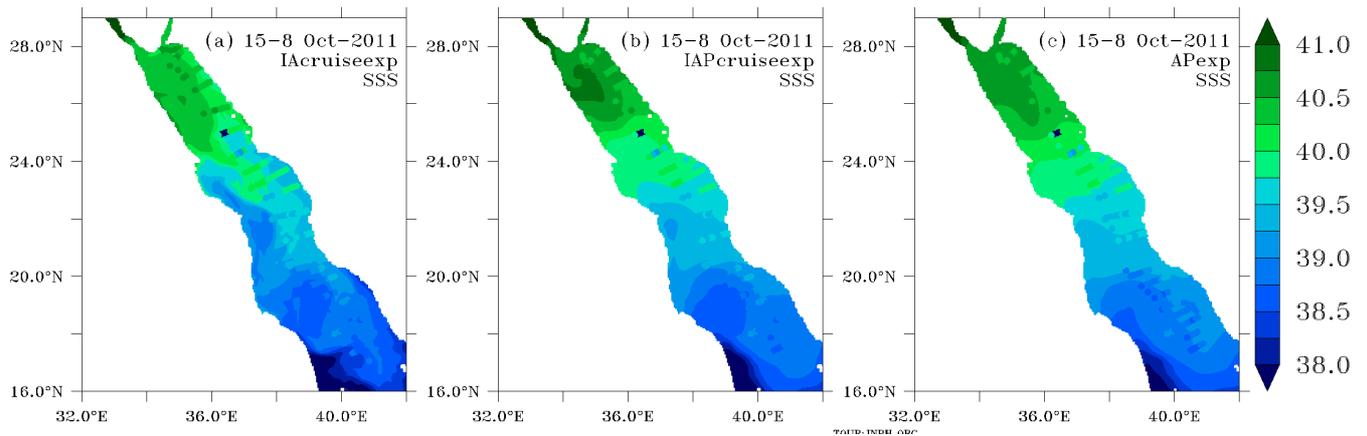

**Figure 11.** Spatial maps of temporally averaged SSS (psu) from (a) *IAcruiseexp*, (b) *IAPcruiseexp,* and (c) *APexp* during the WHOI/KAUST summer cruise period (15[th] September -8[th] October, 2011). Near surface in situ salinities from the CTD data collected during the summer cruise is also layed out with filled circles on each plot.

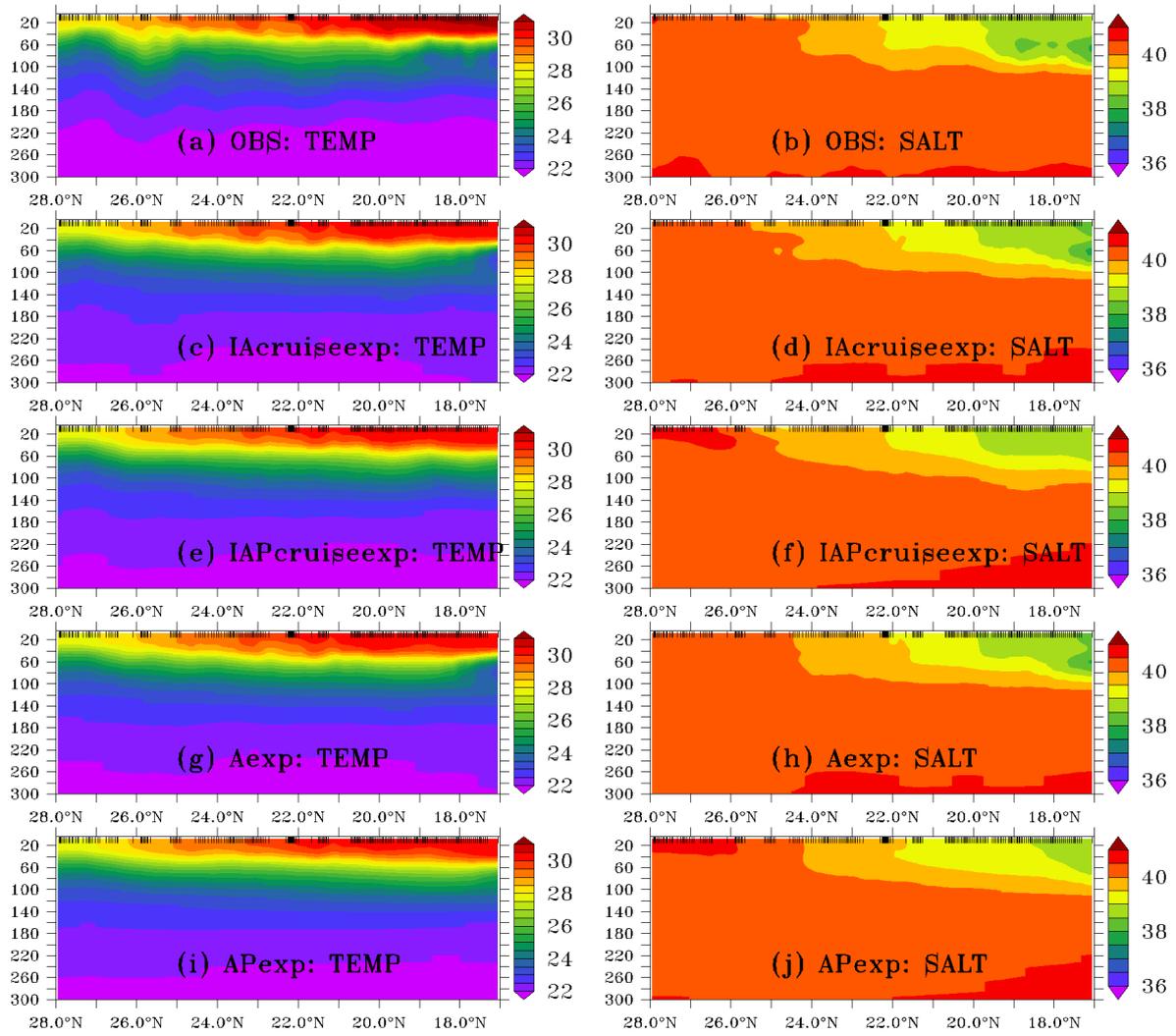

**Figure 12.** Same as Figure 6 except that the results are from *IAcruiseexp* (c-d), *IAPcruiseexp* (e-f), *Aexp* (g-h), and *APexp* (i-j).

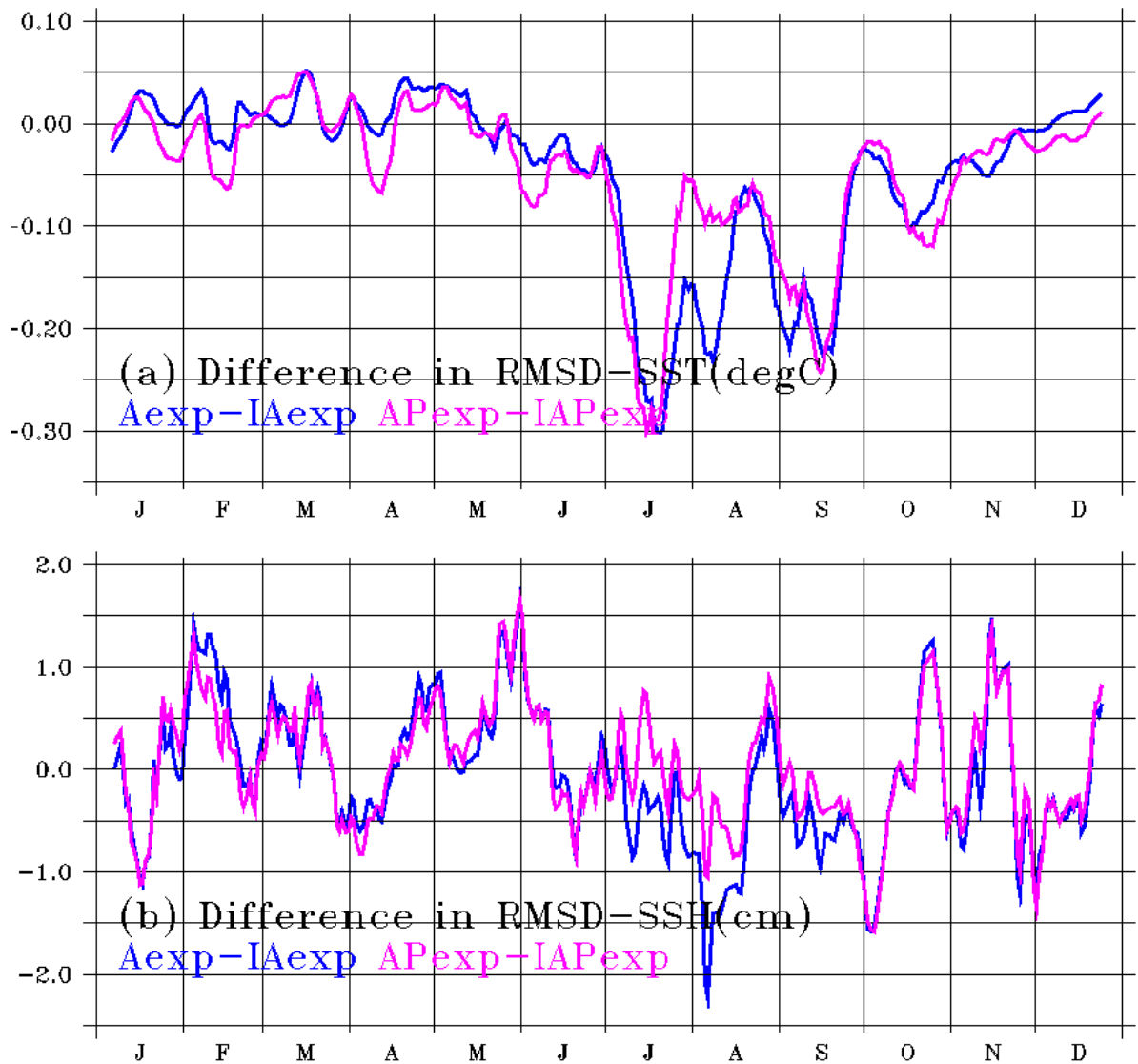

**Figure 13.** (a) Difference between SST RMSDs (°C) of *Aexp* and *IAexp* (*Aexp-IAexp*; blue). Positive value indicate degradation in *Aexp* from *IAexp* and vice versa. The pink line in the figure is for the difference between *APexp* and *IAPexp* (*APexp-IAPexp*). Panel b is ame as "a" but for SSH (cm). RMSD is computed by collocating the daily averaged model forecasts onto the corresponding observation locations. 10-day smoothing is applied to better highlight the differences between the assimilation results.